# Solar Thermoradiative-Photovoltaic Energy Conversion

Eric J. Tervo,[a,1] William A. Callahan,[a,b] Eric S. Toberer,[a,b] Myles A. Steiner,[a] and Andrew J. Ferguson[a]

[a]*National Renewable Energy Laboratory, Golden, CO 80401*
[b]*Department of Physics, Colorado School of Mines, Golden, CO 80401*
[1]eric.tervo@nrel.gov

We propose a solar thermal energy conversion system consisting of a solar absorber, a thermoradiative cell or negative illumination photodiode, and a photovoltaic cell. Because it is a heat engine, this system can also be paired with thermal storage to provide reliable electricity generation. Heat from the solar absorber drives radiative recombination current in the thermoradiative cell, and its emitted light is absorbed by the photovoltaic cell to provide an additional photocurrent. Based on the principle of detailed balance, we calculate a limiting solar conversion efficiency of 85% for fully concentrated sunlight and 45% for one sun with an absorber and single-junction cells of equal areas. Ideal and nonideal solar thermoradiative-photovoltaic systems outperform solar thermophotovoltaic converters for low bandgaps and practical absorber temperatures. Their performance enhancement results from a high tolerance to nonradiative generation/recombination and an ability to minimize radiative heat losses. We show that a realistic device with all major losses could achieve increases in solar conversion efficiency by up to 7.9% (absolute) compared to a solar thermophotovoltaic device under low optical concentration. Our results indicate that these converters could serve as efficient heat engines for low cost single axis tracking systems.

To achieve an electricity grid based on renewable generation, intermittent sources including solar energy must be paired with storage. Thermal energy storage is a very attractive solution due to its simplicity, scalability, and low cost [1-5], especially compared to electrochemical battery storage [6]. However, thermal storage precludes the use of direct solar-to-electricity conversion with photovoltaics (PVs) unless extremely high storage temperatures are used [7]. Instead, sunlight is absorbed as heat and used to immediately or later (with thermal storage) drive a heat engine. Modern concentrating solar power plants accomplish this with thermomechanical cycles that use large turbomachinery, resulting in high capital costs [8, 9]. Accordingly, concentrating solar power plants generally must be very large for cost-competitive electricity generation. This has helped to motivate research into alternative, solid-state heat engines that could also offer simplicity, scalability, and low cost [10-12].

One type of solid-state heat engine that has received significant attention is the thermophotovoltaic (TPV) converter [13-15]. A TPV system consists of a hot emitter of thermal infrared photons which replaces the sun and a PV cell that converts those photons to electricity [16-18]. When the emitter is heated directly or indirectly (via thermal storage) by sunlight, this is a solar TPV system as illustrated in Fig. 1A. Solar TPVs have a very high maximum theoretical solar conversion efficiency of 85% for fully concentrated sunlight on a black absorber [19]. This has motivated a number of theoretical [20-26] and experimental [14, 27-31] studies of solar TPVs, but experimental solar conversion efficiencies have only reached 8.4% [14]. High solar TPV efficiencies are difficult to achieve in practice because they favor relatively high bandgaps (> 0.6 eV) and emitter temperatures (> 1500 K) [11, 19, 21, 32], which also leads to large thermal losses.

Despite their challenges, TPV systems have a number of beneficial characteristics, such as the ability to modify the photon spectrum and recycle unused photons back to the thermal emitter. For example, sub-bandgap parasitic absorption can be drastically reduced by using nanophotonic selective emitters [29, 30, 33, 34] or selectively absorbing cells with a rear mirror [13, 35, 36].

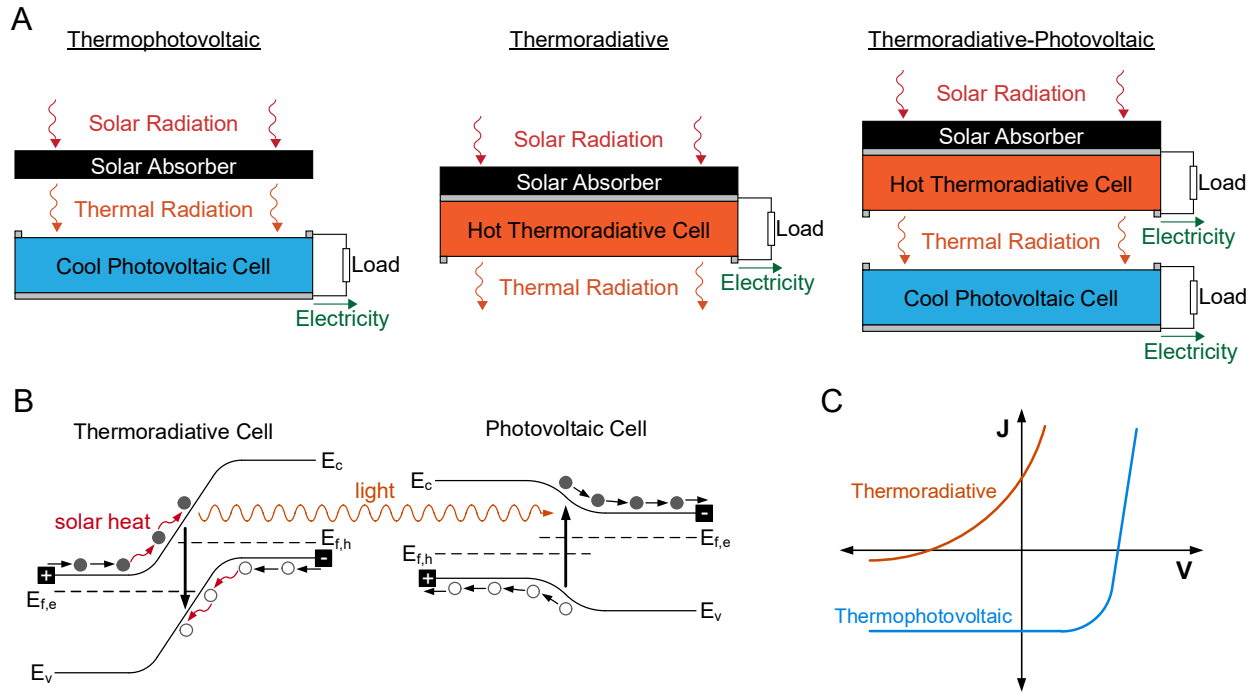

**Fig. 1.** (A) Schematic of a solar thermophotovoltaic, a solar thermoradiative, and a solar thermoradiative-photovoltaic energy converter. (B) Band diagrams of the thermoradiative and photovoltaic cells. (C) Current-voltage diagram of the two devices. A solar thermoradiative-photovoltaic converter produces electricity from both cells simultaneously.

A related technology that could operate efficiently with lower bandgaps and lower hot-side temperatures is the thermoradiative (TR) cell or negative illumination photodiode [37, 38]. TR cells have the same p-n architecture as PV cells, but instead of being illuminated by an external photon source they are directly heated and allowed to thermally radiate to a colder temperature environment, as illustrated in Fig. 1A. The resulting net emission of above-bandgap thermal photons can be thought of as a 'negative illumination' causing a nonequilibrium depletion of minority carriers by radiative recombination. This corresponds to a splitting of quasi-Fermi levels and device voltage opposite that of a PV cell under illumination, as shown in Fig. 1B. A continuous current is enabled by diffusion of charge carriers towards the junction and a sufficient heat supply for those carriers to overcome the junction voltage, which is also illustrated in Fig. 1B. PV cells, on the other hand, rely primarily on drift (movement of charges due to the built-in electric field) to separate electron-hole pairs and enable a continuous current. As a result, TR cells produce power in the second quadrant of a current-voltage plot whereas PV cells produce power in the fourth quadrant, which is depicted in Fig. 1C. Even though TR cells are a relatively new concept, they have already been demonstrated experimentally [39-41] and have been shown to have great potential as emissive energy harvesters [42-49]. As with solar TPVs, TR converters could be used for solar energy conversion by heating the TR cell with sunlight via a solar absorber or thermal storage [50]. However, solar TR systems favor very low bandgap (< 0.3 eV) materials, making them more sensitive to nonradiative losses [38, 50].

To utilize the advantages of both TPV and TR systems, it is natural to consider a heated TR cell emitting to a cool PV cell and obtaining power from both devices [51]. In this paper, we propose such a system for solar energy conversion: a solar TR-PV converter, as shown in Fig 1. We develop a detailed-balance model of the system and use this model to derive its efficiency limit of 85% under maximum concentration. We then consider a more practical configuration of an ideal one-sun area-matched absorber, TR cell, and PV cell, and we show that its efficiency can reach 45% and exceeds that of a solar TR or solar TPV converter for low to moderate bandgaps and absorber temperatures. Finally, we examine a more realistic converter with combined loss mechanisms under optical concentration and demonstrate that it can achieve efficiency gains by up to 7.9% in comparison to solar TPVs under low solar concentration.

## Theory

Let us consider a solar TR-PV system with the energy fluxes illustrated in Fig. 2. A spectral solar radiation flux $q_{sol}(E)$ is incident on the absorber, which may be concentrated or unconcentrated light. For simplicity and clarity when examining performance trends, we model solar radiation as normally-incident light from a blackbody at temperature $T = T_s = 6000$ K, which gives similar efficiency results as the use of the AM 1.5G or concentrated AM 1.5D spectrum [52]. In this framework, the spectral solar radiation flux is given by [19, 53]

$$q_{sol}(E) = \eta_o C_o f_s q_{bb}(E, T_s) \qquad (1)$$

where $E$ is the photon energy, $\eta_o$ is the optical concentration efficiency which we take as unity, $C_o$ is the optical concentration ratio, $f_s = \Omega_s/\pi$ is a geometric factor which accounts for the limited solid angle occupied by the sun, where $\Omega_s = 6.8 \times 10^{-5}$ sr [19, 52, 53], and $q_{bb}(E, T)$ is the spectral blackbody emissive power defined as

$$q_{bb}(E, T) = \frac{2\pi}{h^3 c^2} \frac{E^3}{e^{\frac{E}{kT}} - 1} \qquad (2)$$

Some of this light will be reflected according to the absorber's spectral reflectance $\rho_a(E)$, and the remainder is absorbed assuming that no light is transmitted. Some of the absorbed energy may be lost by emission of thermal radiation, convection, or conduction. Spectral emission losses are given by $\varepsilon_a(E)q_{bb}(E, T_a)$, where $\varepsilon_a(E)$ is the spectral emittance of the absorber. For an opaque surface, $\varepsilon_a(E)$ is equal to the spectral absorptance $\alpha_a(E) = 1 - \rho_a(E)$ according to Kirchhoff's law. $T_a$ here is the absorber temperature and the TR temperature. We group convection and conduction losses according to $h_L(T_a - T_0)$, where $h_L$ is the loss coefficient and $T_0 = 300$ K is the ambient temperature and PV temperature. All energy not reflected nor lost is transferred to the TR cell. In practice, this transfer can be achieved by thermal conduction with direct contact between the absorber and TR cell or through intermediate heat exchangers and a heat transfer fluid as is done in most concentrating solar power plants [8]. With the latter approach, thermal storage may also be integrated into the system in a straightforward manner.

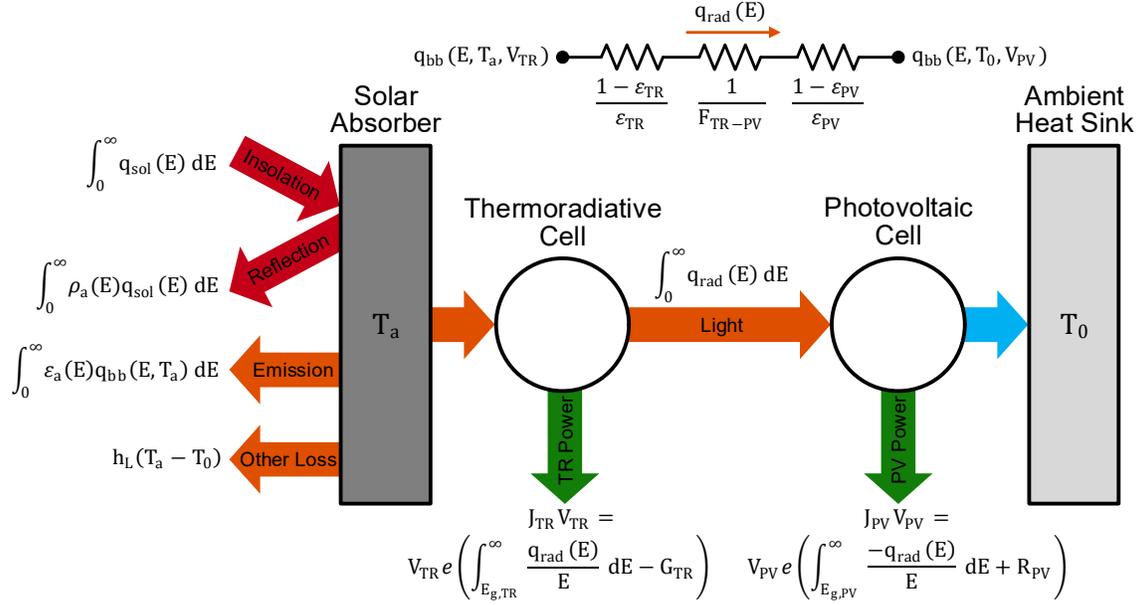

**Fig. 2.** Diagram of energy fluxes and governing equations for the solar thermoradiative-photovoltaic system. The photovoltaic cell is at temperature $T_0$ by thermal coupling to the ambient heat sink, and the thermoradiative cell and absorber are at a temperature $T_a$ determined by an energy balance. The spectral radiative heat flux between the cells, $q_{rad}(E)$, is determined with a series of radiative thermal resistances shown above the schematic and is calculated according to Eq. (7).

The TR and PV cells can be readily modeled with the detailed balance formalism [38, 45, 47] common to PV analysis [53]. For the TR cell, emission of a single above-bandgap photon corresponds to a single charge carrier which may complete a circuit consisting of the TR cell and external load, as illustrated in Fig. 1B. Losses in PV cells are recombination losses, which decrease the population of excess minority carriers towards its equilibrium value (Fig. 1B). In TR cells with a depletion of minority carriers, however, losses are generation losses that increase the population of minority carriers towards its equilibrium value (Fig. 1B). These generation losses include radiative, Auger, Shockley-Read-Hall, and surface processes analogous to those in PV cells [54] which can impede transport and reduce the negative illumination current. The current density in the TR cell can therefore be described by

$$J_{TR} = e\left[\int_{E_{g,TR}}^{\infty} \frac{q_{rad}(E)}{E}\, dE - G_{TR}\right] \qquad (3)$$

where $e$ is the electron charge, $E_{g,TR}$ is the bandgap of the TR cell, $q_{rad}(E)$ is the net spectral radiation flux from the TR cell to the PV cell, and $G_{TR}$ is the nonradiative generation rate. Because $q_{rad}(E)$ is a net quantity, radiative generation losses are included in this expression of the current. Similarly, the current density in the PV cell is

$$J_{PV} = e\left[\int_{E_{g,PV}}^{\infty} \frac{-q_{rad}(E)}{E} + R_{PV}\right] \qquad (4)$$

where $E_{g,PV}$ is the bandgap of the PV cell and $R_{PV}$ is the nonradiative recombination rate. Values for $G_{TR}$ and $R_{PV}$ are determined according to the framework established by Shockley and Queisser [53]. For both the TR cell and PV cell, a factor $f_c$ is defined which represents the radiative fraction of generation/recombination in a p-n diode, such that $f_c = 1$ corresponds to zero nonradiative loss, $f_c = 0.1$ corresponds to 10% radiative and 90% nonradiative generation/recombination, etc. Since the radiative generation/recombination rate is embedded in the $q_{rad}(E)$ term, lower values of $f_c$ increase nonradiative losses without changing radiative generation/recombination, and the voltage dependence of $q_{rad}(E)$ is passed on to $G_{TR}$ and $R_{PV}$. $f_c$ is used to calculate nonradiative loss rates at the ambient temperature $T_0$ for both devices, because this will provide consistent nonradiative losses in the TR and PV cells, and because radiative generation losses in the TR cell result from thermal radiation coming from the PV cell at temperature $T_0$. Any temperature dependence of the cell bandgaps, nonradiative losses, or emittances is neglected.

The net spectral radiative heat flux from the TR cell to the PV cell, $q_{rad}(E)$, in Eqs. (3) and (4) depends not only on the device temperatures and optical properties, but also on their voltages due to luminescence effects. This can be included by modifying the blackbody emissive power as [55, 56]

$$q_{bb}(E,T,V) = \frac{2\pi}{h^3c^2} \frac{E^3}{e^{\frac{E-\mu}{kT}} - 1} \qquad (5)$$

where $\mu = qV$ for $E \geq E_g$ and $\mu = 0$ for $E < E_g$. Because both the TR and PV cells are emitting radiation and may reflect some of that radiation back to the other device, $q_{rad}(E)$ is most easily determined from a radiation resistance network as shown at the top of Fig. 2. The blackbody emissive powers of the TR and PV cells given by Eq. (5) are the two boundary nodes in the network, and the three resistances in series between them are the TR surface resistance (which accounts for the TR cell's optical properties), the space resistance (which accounts for the view factor between the TR and PV cells), and the PV surface resistance (which accounts for the PV cell's optical properties) defined as [57]

$$R_{surf,TR} = \frac{1 - \varepsilon_{TR}(E)}{A_{TR}\varepsilon_{TR}(E)}, \qquad R_{space} = \frac{1}{A_{TR}F_{TR-PV}}, \qquad R_{surf,PV} = \frac{1 - \varepsilon_{PV}(E)}{A_{PV}\varepsilon_{PV}(E)} \qquad (6)$$

where $\varepsilon_{TR}(E)$ and $\varepsilon_{PV}(E)$ are the spectral emittances of the TR and PV cells, $A_{TR}$ and $A_{PV}$ are their surface areas, and $F_{TR-PV}$ is the view factor of the PV cell from the TR cell. We assume that the cells have equal areas and are spaced very closely in comparison to their lateral dimensions, such that $F_{TR-P} \approx 1$ and the area terms may be removed to obtain a radiation flux. The net spectral radiative flux from the TR cell to the PV cell is then calculated as

$$q_{rad}(E) = \frac{q_{bb}(E,T_a,V_{TR}) - q_{bb}(E,T_0,V_{PV})}{R_{surf,TR} + R_{space} + R_{surf,PV}} \qquad (7)$$

The preceding paragraphs fully describe the energy flows in this solar TR-PV system. Once the optical properties ($\varepsilon_a$, $\varepsilon_{TR}$, $\varepsilon_{PV}$, $E_{g,TR}$, $E_{g,PV}$), loss coefficient ($h_L$), voltages ($V_{TR}$, $V_{PV}$) and nonradiative losses ($G_{TR}$, $R_{PV}$) are selected, an energy balance on the absorber and TR cell can be solved for $T_a$. Once $T_a$ is known, it can be used to calculate other performance indicators such as output powers and losses. The solar TR-PV efficiency can then be calculated as

$$\eta = \frac{-A_{TR}J_{TR}V_{TR} - A_{PV}J_{PV}V_{PV}}{A_a \int_0^\infty q_{sol}(E)dE} \tag{8}$$

where $A_a$ is the area of the solar absorber and the negative signs are used because the $JV$ product for both cells is negative. We note that while the discussion thus far has focused on energy flows per unit area, $A_a$, $A_{TR}$, and $A_{PV}$ need not be equal. In fact, there may be advantages in having $A_a < A_{TR}$, because losses from the absorber scale with its area and light can be concentrated to a smaller absorber, and because the output powers of the TR and PV cells scale with their areas [19, 20]. If the areas of the TR and PV cells are not equal, then these cannot be removed from Eq. (6) and the radiation from the TR to PV cell should be described as a total power instead of as a flux.

**Efficiency Limit**

We now turn our attention to the efficiency limit of a solar TR-PV system. The solar conversion efficiency can be broken down into an absorber efficiency $\eta_{abs}$ and a TR-PV efficiency $\eta_{TRPV}$, with $\eta = \eta_{abs}\eta_{TRPV}$. The heat-to-electricity efficiency limit of a TR converter or PV converter alone has been shown to be equal to the Carnot limit [19, 39] under the following conditions: (1) the cell operates in the narrowband limit (emission to the PV cell or from the TR cell occurs only at the bandgap energy), such that sub-bandgap parasitic emission/absorption and thermal losses of above-bandgap photons approach zero; (2) the cell operates in the radiative limit, where nonradiative generation/recombination rates approach zero; and (3) the cell voltage approaches the open-circuit voltage, where the work extracted from each photon is maximized. Applying these conditions to Eqs. (3) and (4) shows us that $q_{rad}(E_g)$ must approach zero, and applying this to Eq. (7) yields $q_{rad}(E_g, T_a, V_{oc,TR}) = q_{rad}(E_g, T_0, V_{oc,PV})$ for an area-matched TR-PV device with aligned bandgaps. Aligned bandgaps are optimal because $E_{g,TR} > E_{g,PV}$ would introduce thermalization losses in the PV cell and $E_{g,TR} < E_{g,PV}$ would result in no PV power in the narrowband limit and would decrease the power density in a real system. $V_{oc,TR}$ and $V_{oc,PV}$ are the open-circuit voltages of the TR and PV cells. Simplifying, we obtain

$$\frac{E_g - eV_{oc,TR}}{T_a} = \frac{E_g - eV_{oc,PV}}{T_0} \tag{9}$$

The corresponding TR-PV efficiency is the ratio of the work extracted per photon (voltage multiplied by electron charge for each device) to the energy supplied to the TR cell per photon (bandgap energy plus potential associated with TR voltage):

$$\eta_{TRPV,lim} = \frac{-eV_{oc,TR} + eV_{oc,PV}}{-eV_{oc,TR} + E_g} \tag{10}$$

When considering Eqs. (9) and (10) it is important to remember that the PV voltage is positive and the TR voltage is negative. Combining and manipulating these equations provides

$$\eta_{TRPV,lim} = 1 - \frac{T_0}{T_a} \tag{11}$$

which is the Carnot efficiency. Although unsurprising, this result is nonetheless instructive as it indicates that a solar TR-PV converter will have the same limiting efficiencies as a solar TR or solar TPV device.

To demonstrate the equivalent limiting efficiencies between solar TPVs and solar TR-PVs, we consider the case of a blackbody absorber. The limiting absorber efficiency is obtained in the absence of conductive and convective losses and is the ratio of the heat received by the TR cell to the incident solar radiation. This efficiency may be determined from Eqs. (1) and (2) and is

$$\eta_{abs,black} = 1 - \frac{T_a^4}{C_o f_s T_s^4} \tag{12}$$

The limiting efficiency for a TR-PV system with a blackbody absorber is therefore

$$\eta_{lim} = \left(1 - \frac{T_a^4}{C_o f_s T_s^4}\right)\left(1 - \frac{T_0}{T_a}\right) \tag{13}$$

Using Eq. (13) to sweep $T_a$ with the maximum possible concentration ($C_o = f_s^{-1} = 4.62 \times 10^4$ suns), one can easily find that this efficiency limit for a solar TR-PV system with a blackbody absorber is $\eta_{lim} = 85\%$ at $T_a = 2544$ K. This matches the maximum efficiency of a solar TPV system [19].

Although 85% is a very high limiting efficiency, this is not a target that can be approached in practice for several reasons. First, the maximum solar concentration and high absorber temperature are not achievable in a real system. Second, this would require an infinitely large TR area due to the narrowband emission approximation, and real cells have an optimum nonzero bandwidth [20, 47, 58]. Finally, nonradiative losses cannot be completely avoided in real materials.

## Results and Discussion

### Ideal One-Sun System

To understand the performance and potential of a solar TR-PV system, we consider a more practical idealized case: no solar concentration, an absorber area equal to the TR area, equal TR and PV bandgaps, and a step-function emittance for the absorber, TR cell, and PV cell, where $\varepsilon = 1$ above the cutoff/bandgap energy and $\varepsilon = 0$ below. As described previously, a smaller absorber area than TR area may be beneficial, but there are fabrication advantages to keeping their areas equal. Matched bandgaps in the TR and PV cells are optimal because this aids in spectral matching of the emission and absorption profiles between the two cells. The effects of conduction/convection losses, nonradiative losses, and optical concentration are neglected here but considered in the following sections. For this system, the spectral radiation fluxes are illustrated in Fig. 3.

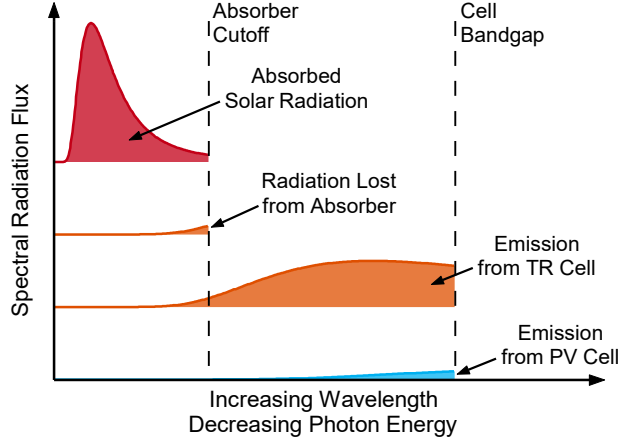

**Fig. 3.** Schematic of spectral radiation fluxes in an ideal solar thermoradiative-photovoltaic system. The red color represents solar radiation from the sun at $T_s$, orange represents radiation from the absorber or TR cell at $T_a$, and blue represents radiation from the PV cell at $T_0$. The cutoff energy of the absorber balances unabsorbed solar radiation with radiation lost from the absorber, and the cell bandgap balances output power with loss due to excess photon energies.

The absorber cutoff energy $E_{abs}$ should balance the unabsorbed (reflected) solar radiation with the lost (emitted) thermal radiation from the absorber. Ideally, this cutoff is located at the intersection of the solar and absorber spectral radiation fluxes. A great deal of research has gone toward the development of selective solar absorbers, which can now achieve solar absorptances in excess of 95% and infrared emittances less than 5% [8, 59, 60]. Recent work on novel materials such as plasmonic nanoparticles [61] or hyperbolic metamaterials [62] have shown the potential for further improvement. The cell bandgap should be low enough that substantial thermal emission is above its energy but high enough to prevent significant thermal losses of above-bandgap photons. Selective emission and absorption by semiconductor p-n diodes can be achieved by multiple strategies [13, 29, 30, 33, 35, 36] as described in the introduction. Of these approaches, thin-film cells with back reflectors [13, 36] are particularly promising due to their relatively simple design and potential for extremely high sub-bandgap reflectance.

With these idealized inputs, we calculate the efficiency of the one-sun area-matched solar TR-PV system as a function of the cell bandgap energy and show the results in Fig. 4A. For comparison, the efficiency of a solar TPV system ($V_{TR} = 0$) and of a solar TR system ($V_{PV} = 0$) are also shown. We use a Nelder-Mead simplex algorithm [63] implemented in MATLAB to find the optimum $V_{TR}$, $V_{PV}$, and $E_{abs}$ for each bandgap energy and solve for $T_a$, $J_{TR}$, $J_{PV}$, and $\eta$. The corresponding absorber cutoff energies and absorber temperatures for the solar TR-PV, TPV, and TR systems are plotted in Fig. 4B.

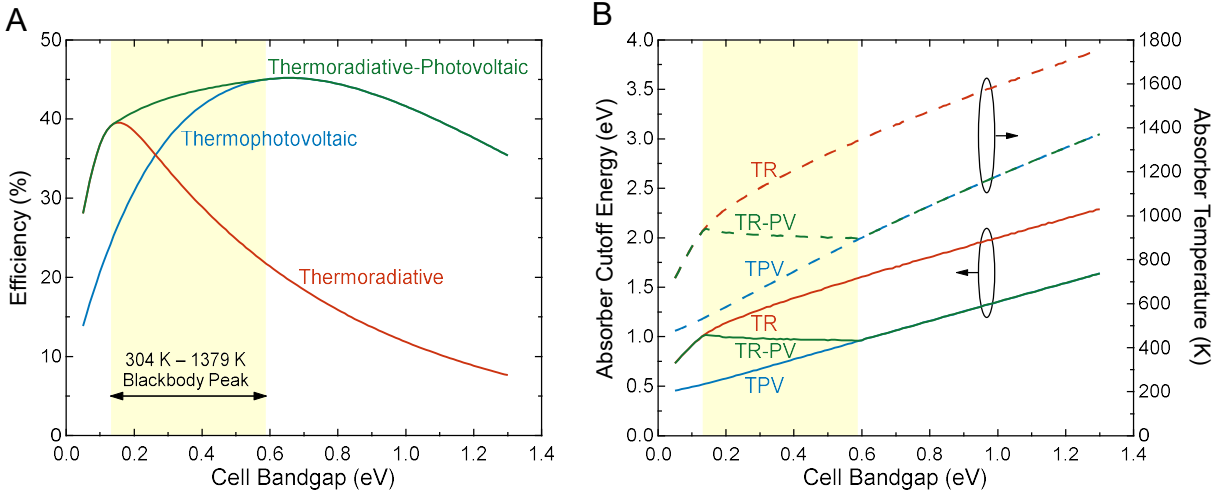

**Fig. 4.** (A) Efficiency limits of ideal one-sun solar thermoradiative, solar thermophotovoltaic, and combined solar thermoradiative-photovoltaic devices as a function of the cell bandgap energy. The thermoradiative-photovoltaic system outperforms the other two devices in the highlighted region of bandgap energies, which corresponds to peak thermal emission from practically achievable temperatures. (B) Solar absorber cutoff energies and absorber temperatures for the devices in (A).

All three devices can reach impressive efficiencies of about 40% – 45%, despite the use of unconcentrated light and area-matched components. Interestingly, the solar TR-PV device outperforms solar TR and solar PV systems for bandgaps from 0.13 to 0.59 eV, as indicated by the shaded yellow region in Fig. 4A and 4B. This range of bandgaps is particularly important for solar-thermal energy conversion, because lower bandgaps enable higher power densities [58] and allow for a lower temperature absorber. This latter advantage of low bandgaps is evident from the absorber temperatures in Fig. 4B; in the region where TR-PVs outperform the other devices, the optimum absorber temperature is relatively constant at about 920 K. This is much lower than typical solar TPV emitter temperatures used with higher bandgap cells [14, 27, 28, 31], which provides substantial engineering and design advantages. We also see from Fig. 4B that in this region the ideal TR-PV device has absorber temperatures and absorber cutoff energies between those of the solar TR and solar TPV devices. This arises from negative luminescent effects on the radiation exchange. As indicated by Eq. (5), the negative TR voltage reduces above-bandgap emission, leading to a higher absorber temperature and higher optimum absorber cutoff energy to reduce absorber emission losses. From 0.13 to 0.59 eV, $V_{TR}$ is smaller in magnitude for the TR-PV device than it is for the TR device, causing the intermediate values of $E_{abs}$ and $T_a$.

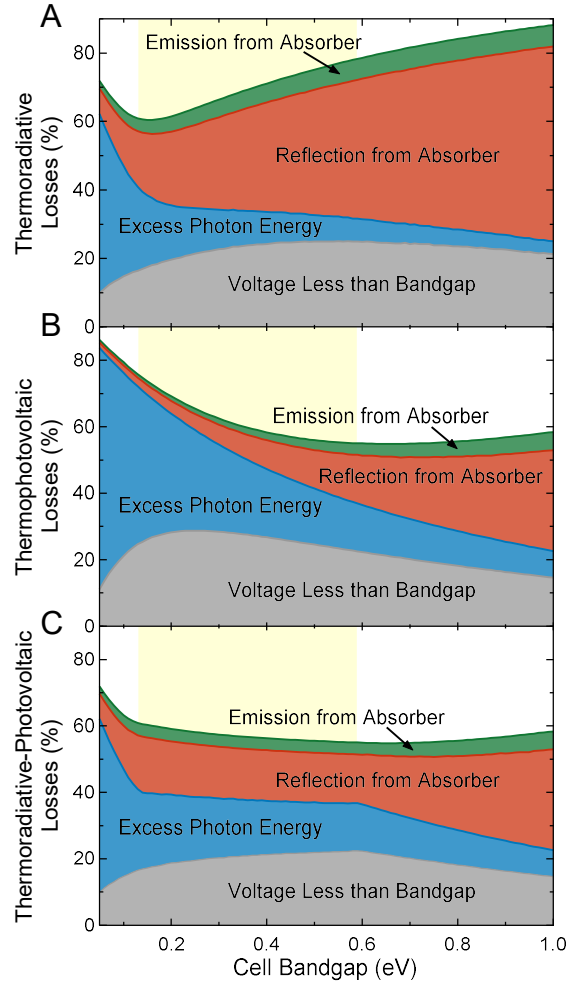

**Fig. 5.** Losses in ideal one-sun (A) solar thermoradiative, (B) solar thermophotovoltaic, and (C) solar thermoradiative-photovoltaic systems corresponding to the devices in Fig. 4. The thermoradiative-photovoltaic system outperforms single devices by optimizing the tradeoff between excess photon energy losses and reflection losses.

The reasons for improved TR-PV performance compared to the other devices are best understood by comparing the different loss mechanisms. For this idealized case, the four losses are emission from the absorber, reflection of solar radiation from the absorber, photon energy exchange between the TR and PV cells in excess of the bandgap energy, and intrinsic losses due to operational voltages less than the bandgap. These losses are depicted for the TR, TPV, and TR-PV devices in Fig. 5A, B, and C, respectively. For all three systems, emission from the absorber is a fairly small loss mechanism. More significant are reflection from the absorber and excess photon energy losses. Reflection losses are higher for the TR device in Fig. 5A than for the TPV device in Fig. 5B because the negative TR voltage suppresses its thermal emission. This leads to a higher absorber temperature and higher optimal absorber cutoff energy (see Fig. 4B). On the other hand, excess photon energy losses are lower for the TR device in Fig. 5A than for the TPV device in Fig. 5B for a similar reason; the negative TR voltage suppresses the thermal emission of photons far above the bandgap. From 0.13 to 0.59 eV, the TR-PV device can balance these two loss mechanisms by utilizing both a TR cell and a PV cell. The voltages of each and absorber cutoff energy can be optimized to reduce the high reflection losses of a TR system while also reducing the high excess photon energy losses of a TPV system, as shown by Fig. 5C. Below this bandgap range the ideal solar TR-PV converter acts only as a solar TR converter ($V_{PV} = 0$), and above this bandgap range the solar TR-PV converter acts only as a solar TPV converter ($V_{TR} = 0$).

## Effects of Non-Idealities

Because real devices cannot avoid conduction, convection, and nonradiative losses, it is important to investigate how these affect system efficiency. Conduction/convection losses are included with a nonzero loss coefficient $h_L$ as shown in Fig. (2). Nonradiative losses are counted by setting the radiative fraction of generation/recombination $f_c$ to values less than one, as described earlier. The effects of conduction/convection losses and nonradiative losses on solar TR-PV efficiency are plotted in Fig. 6A and B, respectively. The green highlighted regions of the curves in each plot indicate the bandgaps for which a solar TR-PV system outperforms a solar TR or solar TPV converter alone, and the temperatures listed indicate the average absorber/TR temperature for these regions, which is approximately constant.

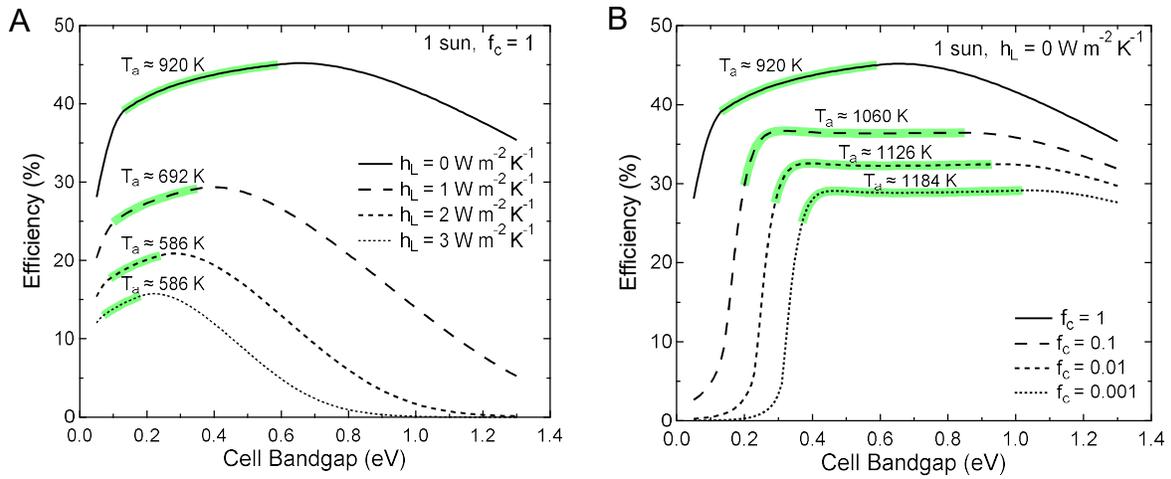

**Fig. 6.** Effects of (A) conductive/convective heat losses from the solar absorber and (B) nonradiative losses on efficiency. The highlighted regions indicate where a solar thermoradiative-photovoltaic device outperforms a solar thermoradiative or solar thermophotovoltaic system alone, and the average (approximately constant) absorber temperature is listed for each of these regions.

Unsurprisingly, conduction/convection losses from the absorber in Fig. 6A tend to decrease solar TR-PV performance. With additional heat losses, the lower absorber temperatures of solar TPV converters lead to reduced efficiency at high bandgaps and a solar TPV efficiency peak at lower bandgaps, as seen in Fig. 6A. This causes the green highlighted region to shift to lower bandgaps as heat losses increase. Because these have a large impact on the efficiency of one-sun TR-PV systems, it is crucial to minimize heat losses in practical systems. A different method to reduce their impact is to operate under optical concentration, which is examined in the following section. Nonradiative losses in Fig. 6B also decrease efficiency as expected, but the TR-PV performance is surprisingly robust in the presence of these losses, even when $f_c$ is very small. Below a bandgap of 0.5 eV, for instance, an $f_c$ value of 0.001 yield large nonradiative loss rates between $10^{25}$ and $10^{29}$ cm$^{-3}$ s$^{-1}$ in the TR cell and between $10^{21}$ and $10^{23}$ cm$^{-3}$ s$^{-1}$ in the PV cell, but the solar conversion efficiency still reaches about 29%. This results from the fact that solar TR-PV devices operate at lower magnitude $V_{TR}$ and $V_{PV}$ than TR or TPV converters alone, and nonradiative losses tend to have an exponential dependence on cell voltage [53]. The combined system is also more tolerant to nonradiative losses in the TR cell,  because no energy is lost to the surroundings when there is a nonradiative generation event. When this occurs, the TR cell current is reduced but the PV cell can still utilize the energy of the emitted photon. As a result of the TR-PV system's ability to retain energy with nonradiative generation in the TR cell, the combined system outperforms the other devices over a larger bandgap range when nonradiative losses are considered.

**Combined Losses with Optical Concentration**

Although examining different loss mechanisms on their own is instructive to understand how they affect performance, it is clear from some of the differing trends in Fig. 6 that their interaction will be significant. Additionally, operating these devices under optical concentration could allow smaller devices to be made at lower cost. We therefore consider a solar TR-PV system with combined losses under varying optical concentration selected to be representative of a practical system.

For the absorber, we model conductive/convective losses with a loss coefficient of $h_L = 1$ W m$^{-2}$ K$^{-1}$, which is a reasonable assumption for evacuated absorbers [8] and is beginning to be approached for non-evacuated absorbers coated with insulating transparent aerogels [64, 65]. Nonideal optical properties are included by setting $\varepsilon_a(E) = 0.98$ above $E_{abs}$ and $\varepsilon_a(E) = 0.02$ below $E_{abs}$, which is similar to materials currently available [8, 59, 60] and is being demonstrated with novel materials [61, 62] as discussed earlier. We choose an absorber cutoff energy of 1 eV based on the results discussed previously. Instead of modeling the incident solar energy as light from a blackbody, we use the AM 1.5D spectrum [66] multiplied by $C_o$ for optical concentration.

For the TR and PV cells, we choose cell bandgaps of 0.35 eV, which is in the range where solar TR-PV converters are expected to outperform the other devices. This is also near the efficiency peaks observed in Fig. 6A and 6B when different loss mechanisms are included. Furthermore, this bandgap is low enough to have substantial thermal radiation above its bandgap. Finally, this bandgap is achievable by epitaxial growth of the InGaAsSb material system [67-69]. Nonradiative generation/recombination is included in the cells by setting $f_c = 0.01$. Nonideal optical properties of the cells are included by setting $\varepsilon_{TR}(E)$ and $\varepsilon_{PV}(E)$ to 0.95 above $E_g$ and 0.02 below $E_g$, which can be achieved by thin-film cells with a good back reflector and front anti-reflectance coating [13, 36].

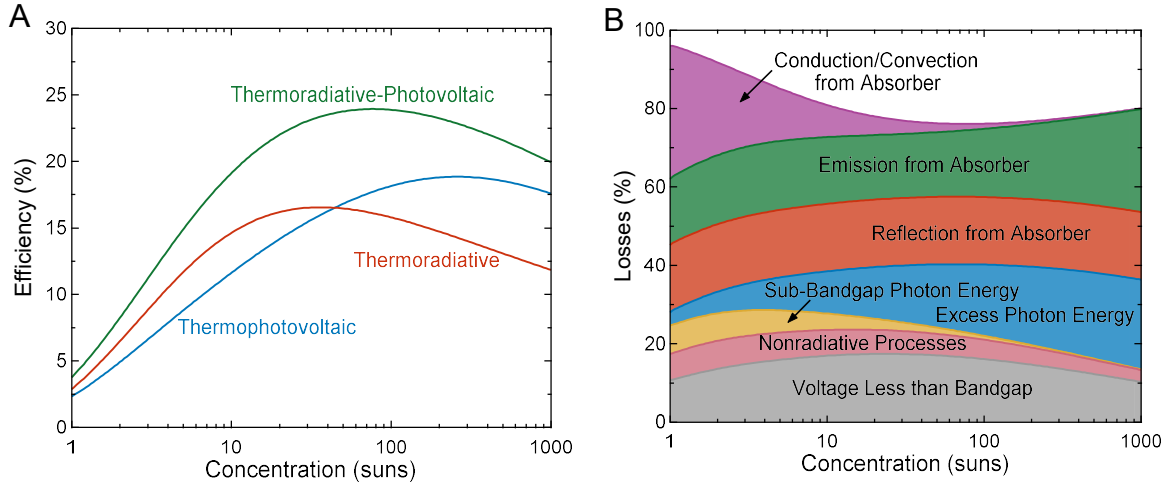

**Fig. 7.** (A) Efficiency of 0.35 eV bandgap solar thermoradiative, solar thermophotovoltaic, and solar thermoradiative-photovoltaic systems with heat losses $h_L = 1$ W m$^{-2}$ K$^{-1}$, nonradiative losses $f_c = 0.01$, and nonideal optical properties for varying solar concentration. (B) Losses for the thermoradiative-photovoltaic system in (A). The remaining white region above the losses represents the useful output power.

With these parameters, we plot the solar TR-PV efficiency in Fig. 7A and the associated loss mechanisms in Fig. 7B as a function of optical concentration. For comparison, the efficiency of a solar TPV and solar TR converter with the same inputs is also shown in Fig. 7A. The solar TR-PV system outperforms the TR and TPV devices for all concentration ratios and achieves a maximum efficiency of about 24% at 80 suns. This is 1.27 times higher than the maximum TPV efficiency and 1.45 times higher than the maximum TR efficiency, representing a substantial performance improvement over the other devices. At lower concentrations, its performance improvement over TPV systems is even more pronounced, reaching up to a 7.9% absolute efficiency gain at 18 suns. It is also worth noting that the peak TR-PV efficiency is similar to that of commercial concentrating solar power plants [8], and the relatively low concentration of the efficiency peak means that solar TR-PV systems could serve as heat engines for low cost single-axis tracking systems. Additionally, this TR-PV converter could likely be optimized to further boost its efficiency, and it could be coupled to a thermal storage system for continuous electricity generation.

The reasons for an optimum concentration of 80 suns with this system can be understood by examining the losses in Fig. 7B. Conduction/convection from the absorber accounts for a large portion of losses at low concentration as expected from the one-sun results in Fig. 6A. These losses scale with $T_a$, but the power produced scales with $T_a^4$, making heat losses less significant as concentration increases and causing the efficiency to rise. As concentration increases further and $T_a$ passes about 1400 K, thermal emission from the absorber and thermal losses of far above-bandgap photons begin to dominate, and the efficiency decreases. We emphasize that the peak efficiency value and the concentration at which it is reached are specific to the combination of losses, bandgap, and absorber cutoff energy selected for this system. These could be tuned to design for a particular concentration ratio, a specific bandgap, or other system parameters.

The breakdown of losses in Fig. 7B can also provide some insight and guidance to further improve solar TR-PV performance. With nonideal optical properties, parasitic sub-bandgap absorption and emission occurs between the TR and PV cells. This decreases at higher concentration when the absorber temperature rises, because more thermal radiation occurs above the bandgap. With improved cell spectral selectivity, this loss can be substantially reduced at low concentration. Emission and reflection losses from the absorber also result in part from nonideal optical components, but these are less sensitive to concentration. Improved spectral selectivity here would reduce these losses across a broad range of concentration ratios. Conduction/convection losses from the absorber display the most drastic dependence on concentration and are most substantial at concentrations less than 10 suns. To enable very low concentration and low temperature solar TR-PV systems, conduction/convection losses must be very low. Finally, we note that increasing the ratio of the TR area to the absorber area could have a substantial impact on these losses and the resulting device efficiency. This could be a promising strategy to minimize conduction, convection, and emission losses from the absorber. We would expect the performance advantages of solar TR-PV cells to extend to these system configurations as well.

## Conclusions

Solar TR-PV systems display clear performance benefits compared to solar TR or solar TPV systems, which results from their ability to utilize advantages of each device. These performance enhancements occur at low to moderate bandgaps in an ideal range for solar-thermal energy harvesting. Similar to a solar TPV or solar TR device, heat losses and nonradiative losses degrade performance, but TR-PV systems can still achieve high efficiencies and outperform individual devices when these are considered. A model of a 0.35 eV bandgap solar TR-PV system with combined loss mechanisms shows that they can substantially outperform solar TPV and solar TR converters at low optical concentrations. Importantly, solar TR-PV converters can be paired with thermal storage to provide reliable electricity generation even with intermittent sunlight.

A solar TR-PV device is necessarily more complex than a solar PV, a solar TPV, or a solar TR converter. This will lead to a higher cost on a per cell area basis than the other devices, which motivates TR-PV use when it can achieve substantially higher efficiencies than the other uncombined devices. For example, the TR-PV system examined in Fig. 7 displays the most significant efficiency gains compared to both a solar TR or solar TPV system at a concentration of about 45 suns. This would be a good operational target for this particular TR-PV device, as it could potentially achieve lower cost per unit power for this concentration. Low levels of concentration such as 45 suns are also not typically targeted for solar TPV energy conversion, because TPVs favor higher temperature emitters and higher bandgaps as shown by Fig. 4. A TR-PV system could therefore enable efficient solar energy conversion for low concentration or in conjunction with hybrid photovoltaic-thermal systems [70], which reduces the need for costly high-performance tracking, optics, and cooling accessories.

Although this preliminary analysis of solar TR-PV systems is encouraging, additional research is needed to model and test their operation in greater detail. In particular, the effect of TR temperature on nonradiative losses could be significant depending on the TR material [47, 71]. Auger losses can increase significantly as temperature increases, which may require the use of strategies for Auger suppression such as the use of a p-i-n diode to reduce carrier concentrations in the active region [72, 73], interface-induced Auger suppression in type-II and type-III heterostructures [74-76], or careful engineering of the confinement potential in nanostructure-based devices [77]. The bandgap of a device also generally shifts with temperature, which could require different material compositions of the TR and PV cells in order to align their bandgaps. Other more practical considerations include investigating the effects of heat losses at the edges of the devices and studying large TR to absorber area ratios, which could boost performance beyond what has been shown here. Nevertheless, this promising initial performance comparison suggests that solar TR-PV systems could be a path to efficient and low-cost solar-thermal energy conversion.

## Acknowledgements

The authors would like to thank Dr. Andreas Pusch of the University of New South Wales for helpful discussions. This work was authored by the National Renewable Energy Laboratory (NREL), operated by Alliance for Sustainable Energy, LLC, for the U.S. Department of Energy (DOE) under Contract No. DE-AC36-08GO28308. This work was supported by the Laboratory Directed Research and Development (LDRD) Program at NREL. The views expressed in the article do not necessarily represent the views of the DOE or the U.S. Government. The U.S. Government retains and the publisher, by accepting the article for publication, acknowledges that the U.S. Government retains a nonexclusive, paid-up, irrevocable, worldwide license to publish or reproduce the published form of this work, or allow others to do so, for U.S. Government purposes.